\documentclass{article}
\usepackage{amsmath,amsfonts,amsthm}
\usepackage{amstext}
\usepackage{amsgen}
\usepackage{amsbsy}
\usepackage{amsopn}
\usepackage{amssymb}
\usepackage{graphicx}
\usepackage{epsfig}
\newtheorem{thm}{Theorem}[section]

\newtheorem{prop}[thm]{Proposition}
\theoremstyle{definition}
\newtheorem{defn}[thm]{Definition}
\theoremstyle{remark}
\newtheorem{rem}[thm]{Remark}
\numberwithin{equation}{section}

\newcommand{\Real}{\mathbb R}
\newcommand{\RealClosed}{\overline{\mathbb{R}}}

\newcommand{\bigO}{\mathbf{O}}

\def\IMAGESPATH{.}

\def\BIBPATH{.}

\begin{document}

\title{The Complexity of Euclidian 2 Dimension Travelling Salesman Problem versus General Assign Problem,
NP is not P}
\author{Carlos Barr\'{o}n Romero \\
cbarron@correo.azc.uam.mx\\ \\
Universidad Aut\'onoma Metropolitana\\
Unidad Azcapotzalco, \\
Divisi\'on de Ciencias B\'asicas e Ingenier\'ia  \\
Departamento de Ciencias B\'asicas\\
Av. San Pablo No. 180, Col. Reynosa Tamaulipas, C.P. 02200, \\
MEXICO }


\date{2010}

\maketitle

\begin{abstract}
This paper presents the differences between two NP problems. It
focuses in the Euclidian 2 Dimension Travelling Salesman Problems
and General Assign Problems. The main results are the triangle
reduction to verify the solution in polynomial time for the former
and for the later the solution to the Noted Conjecture of the
NP-Class, NP is not P.


Algorithms, NP, Numerical optimization.

\end{abstract}
\section{Introduction}

In \cite{ arXiv:Barron2010} there are a complete introduction to
this subject, an analysis in order to verify a solution for the
Euclidian 2 Dimension Travelling Salesman Problem (E2DTSP) and the
General Assign Problem (GAP). A review of the results for the
travelling Salesman Problem is in~\cite{dmj:Applegate1998}, and
for the NP Problems is in~\cite{coe:Woeginger2003}

The next proposition is the key of this research: Given an
arbitrary and large GAP$_n$, it has not an polynomial algorithm
for verifying its solution.

Here a slight different approach is presented to differentiate and
focuses only in the classes E2DTSP and GAP in order to determine
that does nos exist a polynomial time algorithm for solving the
later, i.e., there is not a polynomial time algorithm for solving
the Hard NP problems. The notation, results and a more details are
in \cite{arXiv:Barron2010}.

Section~\ref{sc:GAP_2DTSP} depicts the similarities,
characteristics and propositions of the GAP$_n$ and the
E2DTSP$_n$. The main result is that an arbitrary and large
E2DTSP$_n$ preserves the inhered by construction the triangle
inequality property. On the other hand, GAP$_n$ does not preserve
or inhere any property that could drive toward the construction of
a verifying algorithm of its solution. Therefore, it is impossible
to build an efficient algorithm for solving arbitrary and large
GAP$_n$.The
 last section contains conclusions
and future work.

\section{The General Assign Problem and the Euclidian 2 Dimension Travelling
Salesman Problem}~\label{sc:GAP_2DTSP}

The notation is the same as in~\cite{ arXiv:Barron2010}.
\begin{defn} {Notation}
\begin{enumerate}
    \item GAP$_n$ is the General Assign Problem of size $n$.

    \item E2DTSP$_n$ stands for the version of the Traveller Salesman Problem with $_n$
vertices. They come from points in a plane (as subset of
$\Real^2$) and the weight of the edges are the euclidian distance
between vertices.
\end{enumerate}
\end{defn}

 In this section the GAP$_n$ and E2DTSP$_n$압 characteristics and
 properties are presented. It is intended to give general results
 for TSP$_n$ and for the specific case E2DTSP$_n$.

The GAP$_n$ consists a complete graph, a
function which assigns a value for each
 edge, and objective function,
$\text{GAP} \_n$ $=$
 $(G_n,c,f)$, where $G_n =\left( V_n, A\right) $,  $V_n$  $\subset$ $\mathbb{N}$
 $A$ $=$
 $\left\{ \left( i,j\right) \, | \, i,j \in V\right\}$,
$c : V \times V \rightarrow  \RealClosed$, and $f$ is a real function
for the evaluation of any path of vertices. Solving a GAP means to look
for a minimum or a maximum value of $f$
    over all cycles of $G_n$.
Note that $c(i,i)=\inf$, and $c(\cdot,\cdot)$ can be see as a
matrix of  $\RealClosed^{\,n \times n}.$

\begin{rem} Some properties and characteristics GAP and E2DTSP are:
\begin{enumerate}
    \item The complete property of $G_n$ means that $\forall i, j \in V_n, \exists e=(i,j)\in A$.
    \item $G_n$ is a directed graph, which means that edges are a ordered pair.
    \item A path of vertices is a sequence of vertices of $V_n$.
    \item A cycle is a path of different vertices but the first and the last
    vertex are the same. Hereafter, a complete cycle is a cycle
    containing the $n$ vertices of $V_n$. Also, a complete cycle has $n$
    edges.
    \item A  E2DTSP$_n$ is a case of a GAP$_n$ where
    $c(i,j)=c(j,i)$ (the cost matrix is symmetric) and $c(i,j) \geq 0, \forall i,j$, and finally
    $f$ is the sum of the edges' cost of a cycle. It is a special case where
    $c(i,j) = \hbox{distance}(v_i,v_j)$ $\forall v_i, v_j \in V_n$, the vertex $v_i \in
    \Real^2,\forall i=1,\ldots,n$, and $\hbox{distance}(v_i,v_j)=\|v_i-v_j\|$, is the Euclidian
    distance. Note that for a  E2DTSP$_n$, we are going to refers to a vertex by its
number or by $v_i$, in this case $v_i$ is a point of $\Real^2$.
\item A sequence of vertices is a path and its denoted by $p_k$.
\item The objective function $f$ is also denoted by $c$, and let
$c(p)$ be the cost function of the path $p$, i.e., it is  given by
the summation of edge압 cost of the consecutive pairs of vertices
of $p$.

\end{enumerate}
\end{rem}

The following propositions are easy to prove and they are well
known results (also they are presented in
\cite{arXiv:Barron2010}).

\begin{prop}
\ \newline
\begin{enumerate}
\item Any GAP$_n$ has a solution. This means that exists a
Hamiltonian cycle with minimum cost.

\item Let $f$ be the minimum or maximum edge압 cost as the
objective function. Then the Hamiltonian cycle of the GAP$_n$
which is the solution can be found in polynomial time.

\item Any GAP$_n$ has
\begin{enumerate}
    \item $n(n-1)$ edges.
    \item $(n-1)!$ complete cycles. The complete cycles can be
    enumerated in $n$ different ways. Hereafter, a cycle means a
    complete cycle.
    \item For $n\geq 4$, the maximum number of coincident edges of
    two different complete cycles is (n-3).
\end{enumerate}

\item The Research Space of GAP$_n$ is finite and numerable, and
it has $(n-1)!$ elements and can be enumerated in $n$ different
ways.

\item Let be $[i_k]_{k=1}^{n+1}$ a given cycle  of a GAP$_n$. Then
$\exists$! $m:[1,\ldots,n]\rightarrow [1,\ldots,n]$ such that the
cost of the cycles do not change for an equivalent GAP'$_n$ with
cost function given by $ c \circ  m^{-1} = c(m^{-1}(\cdot))$ and
$[m(i_k)]_{k=1}^{n+1}=[n,n-1,\ldots,1,n]$, which corresponds to
the first cycle of the descent enumeration of the vertices
beginning with $n$.

\item Let be $[i_k]_{k=1}^{n+1}$ the solution cycle  of a GAP$_n$.
Then $\exists$! $m:[1,\ldots,n] \rightarrow [1,\ldots,n]$ such
that $[n,n-1,\ldots,1,n]$ is the solution of the equivalent
GAP$_n$, which corresponds to the first cycle of the descent
enumeration of the vertices beginning with $n$.

\item Let be GAP$_n$ such that $c(i,j)$ is given by the following
matrix
 $$%
\left\{ \begin{array}{ c c c c c}
          & 1                  &     2          & \cdots   & n-1          \\
   n      &                    &  n \cdot 2     & \cdots   & n \cdot n-1  \\
  \vdots  & \vdots             &  \vdots        & \vdots   & \vdots       \\
  n^{n-2} & n^{n-2}\cdot 2     &  \cdots         &          & n^{n-2}\cdot (n-1)  \\
  n^{n-1} &  n^{n-1}\cdot 2    &  \cdots       & n^{n-1}\cdot (n-1) &   \\
\end{array}%
\right\}
$$
then it has a unique solution and all cycles have different cost.

\end{enumerate}

\end{prop}

Important differences between GAP$_n$ and TSP$_n$ are in the
following proposition. Here $c$ denotes de objective function.

\begin{prop}~\label{Prop:GAP_vs TSP}

\begin{enumerate}
\item For E2DTSP$_n$, $c$ is monotonically increasing,i.e, $c(p_1)
\leq c(p_2)$ where the sequence of vertices of $p_1$ is a
subsequence of $p_2$.

\item For GAP$_n$, $c$ is not monotonically increasing.

\item Let $c' \in \Real $, then for any E2DTSP$_n$, $P(c(p_2) \geq
c'| c(p_1) = c')=1$ where $P$ is the probability function, and
$p_1$ is a sub-path of $p_2$.

\item Let $c' \in \Real$, then for GAP$_n$. $P(c(p_2) \leq c' |
c(p_1) = c')>0$ where $P$ is the probability function, and $p_1$
is a sub-path of $p_2$.
\end{enumerate}
\end{prop}

The following propositions characterizes the probability of
searching the solution by a uniform random search. GAP has $n+1$,
($n$ very large) vertices in order to simplify the notation of the
results.

\begin{prop} Given an arbitrary and large GAP$_{n+1}$. Let be
$m=\min \{ c(j) \,|\, j \hbox{\, is a cycle of GAP}_{n+1} \}.$
\begin{enumerate}
\item The probability of selecting a cycle $j$ of GAP$_{n+1}$ is
$P(j)=\frac{1}{n!}$.
\item $\frac{1}{n!} \leq P(j \, | \, c(j) \leq m)$. If
GAP$_{n+1}$has a unique solution, $j$, then $P(j \, | \, c(j) \leq m)=\frac{1}{n!}.$
\item Let be $J$ a set of cycles of GAP$_{n+1}$, and cardinality of
$|J| = n^3, i.e.,  |J| \sim \bigO((n+1)^3)$. Then
$\exists A, 1 < A < 2^3$ such that $\frac{1}{A(n-3)!} \leq P(J)=
\frac{n^3}{n!} \leq \frac{1}{n!}.$ Also, $\frac{(n-3)!-1}{(n-3)!}
\leq P(J^c) \leq \frac{A(n-3)!-1}{A(n-3)!}.$ This means
$P(J)\approx 0,$ and $P(J^c)\approx 1$.
\end{enumerate}
\begin{proof} \ \newline

\begin{enumerate}
\item It follows from the basic properties of complete graph of
GAP$_{n+1}$.
 \item It follows by the previous statement and the
unique $p_j$.
\item It follows from an appropriate approximation
of $\frac{n!}{n^3}$.
\end{enumerate}
\end{proof}
\end{prop}

\begin{figure}
\centerline{ \psfig{figure=\IMAGESPATH/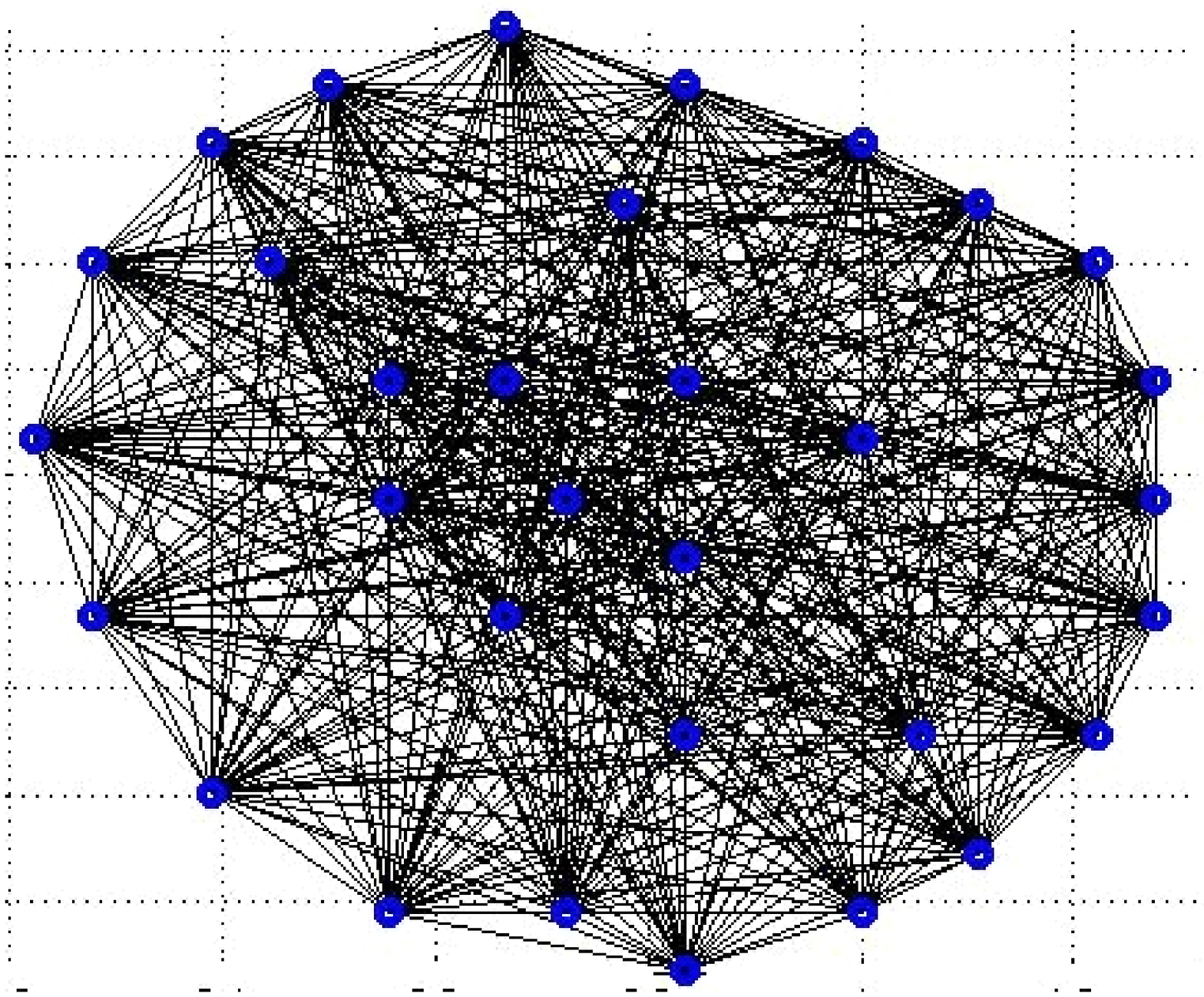,
height=50mm} \psfig{figure=\IMAGESPATH/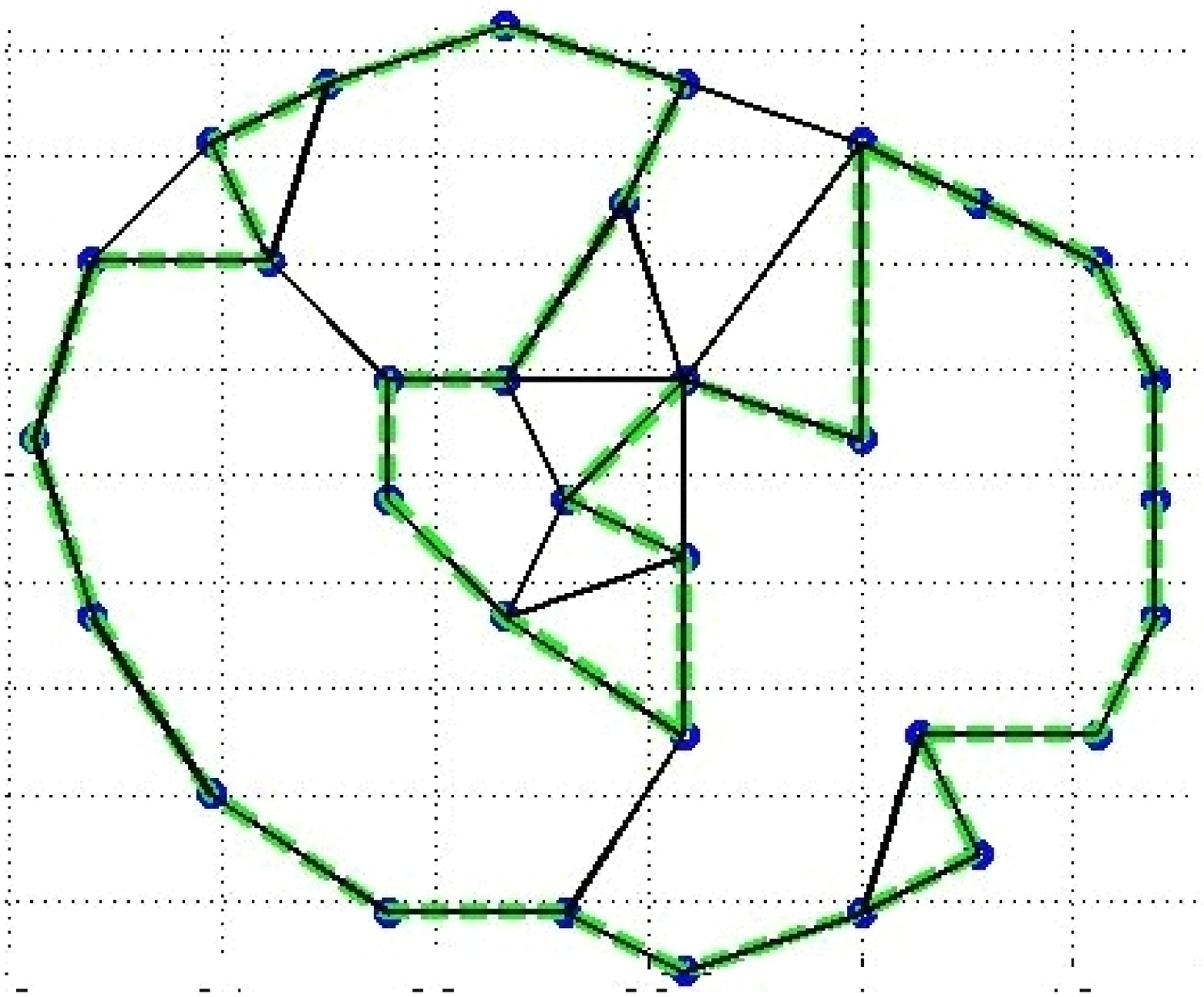,
height=50mm}}
 \caption{Example of a E2DTSP$_n$ Complete and its triangle reduction.}~\label{fig:E2DTSP_completo_reducido}
\end{figure}

\begin{figure}
\centerline{ \psfig{figure=\IMAGESPATH/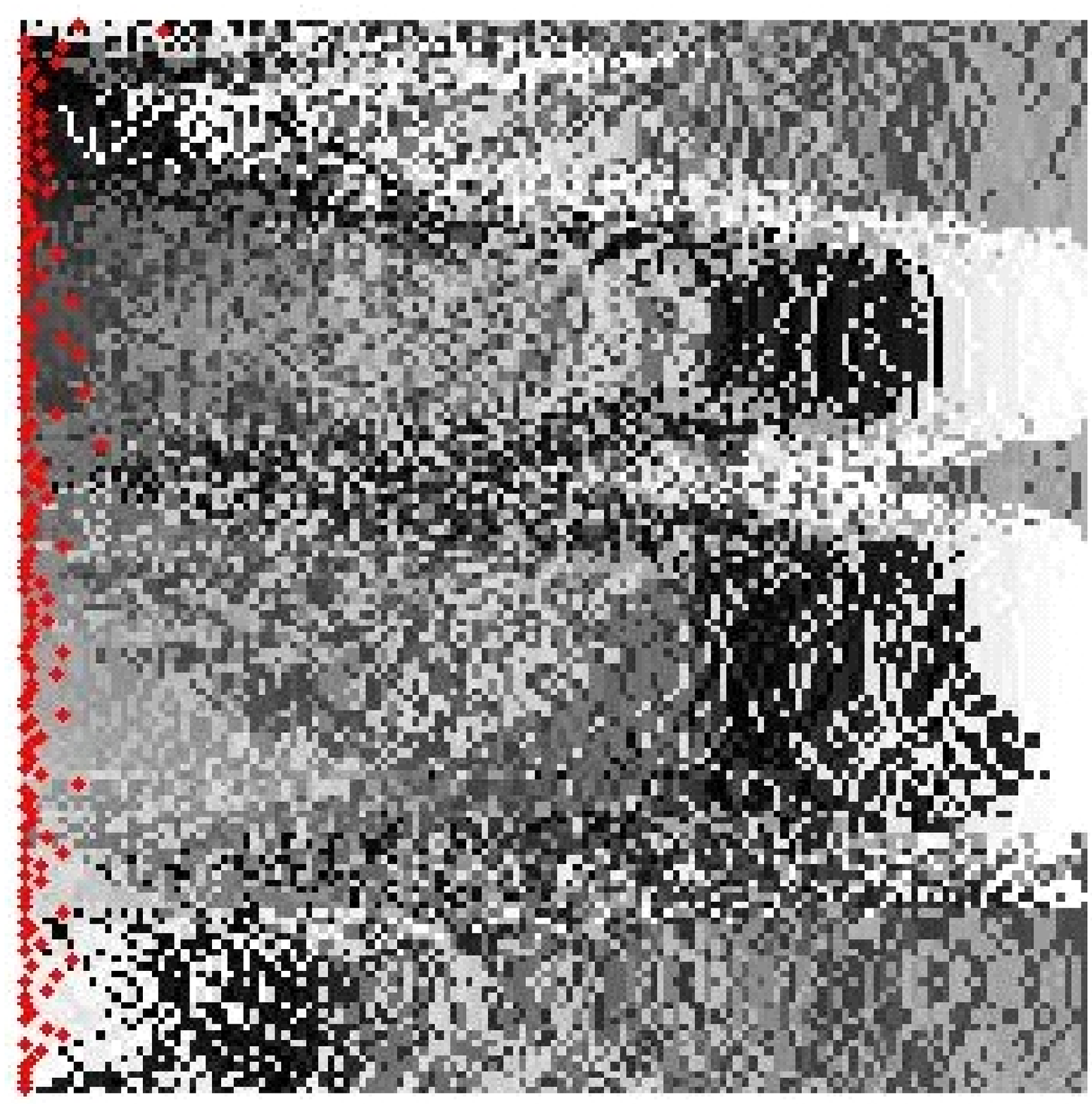,
height=50mm} \psfig{figure=\IMAGESPATH/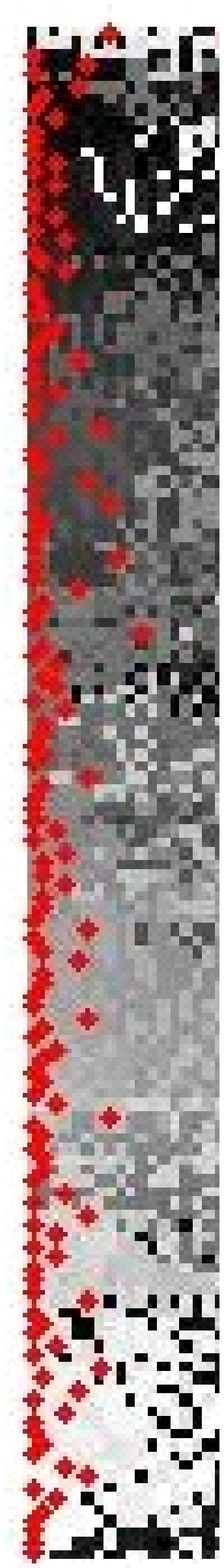,
height=50mm}} \caption{Example of the sorted cost matrix for
E2DTSP$_n$ Complete and its triangle reduction.
}~\label{fig:mapaE2DTSP_completo_reducido}
\end{figure}

\begin{figure}
\centerline{ \psfig{figure=\IMAGESPATH/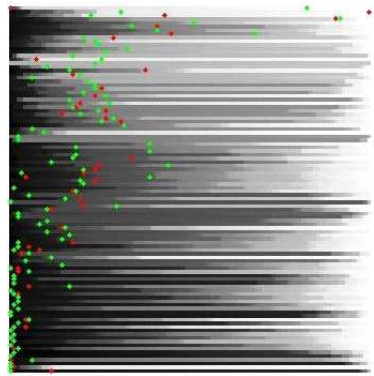,
height=50mm} }
 \caption{Sorted cost matrix for a
 GAP$_n$. The green dots are the solution. The red dots are a very closed solution}~\label{fig:sortedcost_matrix_GAP}
\end{figure}

The last proposition shows that the cycle with minimum cost is
quite difficult to find it, when no other properties are used than
a uniform random search. Moreover, also if the solution is in a
set of Hamiltonian cycles with polynomial cardinality, to look for
it has a probability almost null, and the complementary event is
almost certain. It means that properties are need to trim the
search space in order to estimate or to verify the solution. The
main difference between GAP$_n$ and E2DTSP$_n$ is that the last
one preserve a triangle reducibility property.
In~\cite{arXiv:Barron2010} the reducibility was described for
tubes, here, this property is related triangulation containing the
solution. Fig.~\ref{fig:E2DTSP_completo_reducido} depicts an
example of this cunning property of an E2DTSP$_n$.
Fig.~\ref{fig:mapaE2DTSP_completo_reducido} depicts the
corresponding sorted cost matrix of an of the E2DTSP$_n$. Note
that the triangle reduced sorted cost matrix is elongated,
implying there are very few cycles to verify the solution.

\begin{prop}
Given any E2DTSP$_n$ it is triangle reducible, i.e, it exists a
triangulation of the E2DTSP which preserve the solution.
\begin{proof}
Because the graph압 edge's values of the E2DTSP$_n$ come from
distances between vertices, it is trivial to build a triangulation
for any E2DTSP$_n$ over the Hamiltonian cycle which is the
solution. Moreover, the triangle inequality permits to select the
proper edges to avoid diagonal over the solution. This means that
the reduced E2DTSP$_n$ on this triangulation has the solution of
the original problem by construction.
\end{proof}
\end{prop}

In~\cite{ arXiv:Barron2010} a version of the previous proposition
(\textbf{Prop. 6.9}) was presented to prove the that E2DTSP$_n$
has an algorithm to verify the solution in polynomial time.

The next proposition shows that any E2DTSP$_n$ combined with any
GAP$_m$ loss any property when they are  combined to create a
large GAP$_{n+m}$. Moreover, the class E2DTSP is a proper subset
of the class GAP$_n$, i.e., class E2DTSP $\subset$ class GAP, and
class GAP $\neq$ class E2DTSP.

\begin{prop}
Given a E2DTSP$_n$ (which is triangle reducible) and a GAP$_m$,
then $\exists$ a large arbitrary GAP$_{n+m}$ which is not triangle
reducible.
\begin{proof}
The complete graph of a GAP$_{n+m}$ is builded by inserting
appropriately the graphs of E2DTSP$_n$ and GAP$_m$. The rest of
edges for the $n+m$ vertices can be added with random values.
Finally, the resulting GAP$_{n+m}$ is not reducible because many
of its edges았alues are not from 2D distance between vertices but
random values. Moreover, the graph's edge's values of the
GAP$_{n+m}$ do not comply the triangle inequality.
\end{proof}
\end{prop}

In similar way, we can state the following property of lacking a
property to verify the solution in polynomial  time for arbitrary
and large GAP$_n$. Fig~.\ref{fig:sortedcost_matrix_GAP} depicts
the sorted cost matrix of a GAP$_n$. Because of the connectivity
of the vertices with similar edge' cost, there is not a simple
reduction of the GAP$_n$'s graph as in the case of the E2DTSP$_n$.
Note that the fact that the sorted cost matrix is not reduced,
this imply that even with a putative solution there are many
cycles to compare to verify the optimality of the putative
solution.

\begin{prop}~\label{prop:GAP_do_not_keep_Properties}
Let be GAP$_n$, such that has an algorithm to verify the solution
in polynomial time. Given any GAP$_m$, then $\exists$ a large
arbitrary GAP$_{n+m}$ which has not an algorithm to verify the
solution in polynomial time.
\begin{proof}
The complete graph of a GAP$_{n+m}$ is builded by combining
 appropriately the graphs of GAP$_n$ and GAP$_m$. The
rest of edges for the $n+m$ vertices can be added with random
values. The GAP$_n$'s algorithm provides a reduction of the edges
of GAP$_n$ to consider its solution to get a path, but  the
resulting GAP$_{n+m}$ can change the path so the solution of
GAP$_n$ is not a sub path of the solution of GAP$_{n+m}$.

By example, it is possible to join GAP$_n$ with two
GAP$_\frac{m}{2}$, in the way that GAP$_n$ is like a tube. This
case could preserve the solution of GAP$_n$ but it is not a
general case. Tt is easy to see by inspection of the sorted cost
matrix $\textsl{M}$ that the solution of GAP$_n$ which it is
founded by the polynomial time algorithm must comply to have a
elongated reduction with an appropriate enumeration of the
vertices, otherwise the verification of this solution it is not
possible in polynomial time. This means, that only an appropriate
and few edges are needed to consider for GAP$_n$. However, with an
arbitrary joining between GAP$_n$ and GAP$_m$ a lot of edges with
an appropriately random values not necessarily comply with the
property (if it has one) of the verifying algorithm of GAP$_n$. In
the arbitrary case it is possible that  any vertices of GAP$_n$
can have better alternatives by the added edges' connection with
the vertices of GAP$_m$. By alternatives, it is not only in the
sense of a greedy algorithm, here we can have a sequences of
oscillating edges' random values. Therefore, it exists a
GAP$_{n+m}$ such that it nullifies or overrules any property
inhered by the polynomial time algorithm of GAP$_n$ for verifying
the solution in polynomial time of GAP$_{n+m}$ by the addition of
appropriately random values for its edges.
\end{proof}
\end{prop}

The proposition shows that an algorithm could claim and maybe
solve a NP problem but the verification without properties over an
a worst case with arbitrary data can not be demonstrated or seeing
in polynomial time or by inspection of the sorted cost matrix
$\textsl{M}$, and if the research space could be reduced then
input problem is not a arbitrary worst case with many oscillating
cycles. Prop.~\ref{Prop:GAP_vs TSP} states why E2DTSP$_n$'s
solution is easy to verify but GAP$_n$.

\begin{prop} Given an arbitrary and large GAP$_n$, it has not an polynomial
algorithm for verifying its solution.
\begin{proof}
It is immediately from the previous proposition.
\end{proof}
\end{prop}

\section*{Conclusions and future work}~\label{sc:conclusions and future work}

Here the property of reduction was extended to triangle reduction
for E2DTSP$_n$. This means that these are easy to verify the
solution (and it is very possible to solve they in polynomial
time) but arbitrary GAP$_n$ is not. As it stated in~\cite{
SODA:Amenta07} a triangulation algorithms has
$\bigO\left(n^{\frac{m-1}{p}}\right)$, $\forall p \in [2, m-1]$.
Therefore, it is highly possible that it exists an algorithm to
solve arbitrary and large E2DTSP$_n$. However, the
proposition~\ref{prop:GAP_do_not_keep_Properties} shows that
arbitrary and large GAP$_n$, has not an inherited property
allowing to build an efficient algorithm for verifying the
solution, moreover, to solving it. When there is not structure or
properties the simple machine, the finite automata has not an
efficient implementation as it described
in~\cite{arXiv:Barron2010}.

To my best knowledge this approach to verify in polynomial time
the solution is given by the frontier's position on the sorted
cost matrix $\mathcal{M}$. The figures here depicts the
differences between E2DTSP$_n$ and GAP$_n$. The main result is
from proposition~\ref{prop:GAP_do_not_keep_Properties}, which is
the base of the argumentation why P-Class (represented by
E2DTSP$_n$) and the Hard NP-Class (represented by GAP$_n$) have
totally different complexity. The later has not polynomial
complexity in the general or worst case scenario, i.e., for large
and arbitrary GAP$_n$ but exponential.  The result shows that does
not exist a general property that allows to solve in polynomial
time all Hard NP problems.

For the future, no student or colleague have accepted my
invitation to build the efficient algorithm for E2DTSP$_n$ using
my results and by example the article~\cite{ SODA:Amenta07} for
building a triangulation.

Finally, this closing my previous article~\cite{ arXiv:Barron2010}
in order to solve the Noted Conjecture of the NP-Class, and I
hope, it brings a promising theoretical perspective to address the
construction of efficient algorithms for arbitrary NP problems.

\section*{Acknowledgement}

To whom, it wants to share the knowledge.

This work is dedicated to my family, and my friends: Roland
Glowinski, Ioannis Kakadiaris, Alberto SantaMaria, Miguel Samano,
Felipe Monroy, and COGEA.

\bibliographystyle{abbrv}
\bibliography{\BIBPATH/NPComplexity_v13}

\end{document}